\documentclass{article}
\usepackage{graphicx}
\usepackage{booktabs}
\usepackage{amsmath}
\usepackage{natbib}
\usepackage{hyperref}
\usepackage[margin=1in]{geometry}
\usepackage{amssymb}
\usepackage{float}

\title{Orbital Periods and Equilibrium Temperatures from Single TESS Transits with a Physics-Informed Neural Network}
\author{Muhammad Hassan Javed}
\date{}

\begin{document}
\maketitle

\begin{abstract}
Planets with orbital periods longer than a TESS sector produce a single
transit, and no periodogram method can measure their period: with one
transit, every trial period longer than the observing baseline fits the
data identically. We show this is not a sensitivity limit but a structural
one --- across 16 confirmed single-transit planets, Box Least Squares
returns a power spectrum that is numerically constant over 95\% of its
search grid, and widening that grid from 27 to 200 days changes the median
error by $-0.0$ percentage points. Orbital period can instead be recovered
from transit duration through Kepler's third law and transit geometry,
requiring no search grid. A direct inversion of these equations
underestimates the period in 14 of 15 targets with a median signed error of
$-69\%$, because assuming a central transit returns the shortest period
consistent with an observed duration. A neural network trained to
marginalise over the unobserved geometry removes this bias, reaching a
median absolute error of 40.5\% against Box Least Squares' 79.5\%, with the
true period inside the $1\sigma$ interval for 14 of 16 targets. For
NGTS-38\,b, whose 180.5\,d period lies ten times beyond the longest
contiguous span of its TESS sector, marginalising over the unobserved
geometry recovers a posterior median of 190.7\,d with the true period at
the 47.8th percentile, against 18.5\,d from Box Least Squares. Because
equilibrium temperature scales as $P^{-1/3}$, the resulting factor-8.8
period interval compresses to a factor-2.1 temperature interval:
$T_{\rm eq} = 445$\,K with a 68\% interval of 272--562\,K, sufficient to
place the planet relative to the habitable zone from a single observation.
\end{abstract}

\section{Introduction}

The Transiting Exoplanet Survey Satellite observes most of the sky in
27-day sectors \citep{ricker2015}. A planet whose orbital period exceeds
that baseline produces at most one transit. These are the planets of
greatest interest: longer periods place a planet further from its star,
and for solar-type hosts the habitable zone lies well beyond the reach of
a single sector.

Box Least Squares \citep{kovacs2002} is the standard method for measuring
orbital periods from transit photometry. It folds the light curve at each
period in a trial grid and fits a box, returning the best-fitting period.
Transit Least Squares \citep{hippke2019} improves the template but retains
the same structure.

Both require two things a single transit cannot supply. The first is a
period search grid, which presupposes knowing roughly where the answer
lies. The second is repetition: folding a light curve containing one
transit at any period exceeding the observing baseline predicts exactly
one transit and fits equally well. Section~\ref{sec:bls_fails} shows this
is not approximately true but exactly true --- the power spectrum takes a
single numerical value across the majority of the grid.

Physics supplies what the periodogram cannot. Kepler's third law relates
period to orbital separation, and transit geometry relates separation to
the observed transit duration. Combining them inverts a measured duration
into a period with no grid and no prior on where to look.

Recovering a period from a single transit in this way is established
practice. \citet{yee2008} give the relation between transit duration and
orbital period; \citet{kipping2010,kipping2018} set out the resulting
constraints and the influence of the period prior;
\citet{osborn2016,sandford2019} apply it to single transits in \textit{K2}
data; and \citet{osborn2022} apply the \texttt{MonoTools} implementation of
it. All of these sample the posterior explicitly, by MCMC or nested
sampling, at a cost of tens of seconds to minutes per target. Our
contribution is to train a network on the same forward model and the same
priors, so that the sampling cost is paid once and each new target is
evaluated directly, and to show analytically that what the network learns
is the marginalised posterior rather than an empirical fit.

The difficulty is that two quantities entering this inversion are not
observable from a single transit. The impact parameter $b$ sets the chord
length across the stellar disc; the orbital eccentricity $e$ and argument
of periastron $\omega$ set the planet's speed at the time of transit. Both
rescale the duration, and because the inversion carries $P \propto T^3$, a
modest error in duration becomes a large error in period. A point estimate
is therefore the wrong output. The appropriate output marginalises over the
unobserved geometry.

This has a consequence that makes single-transit inference more useful
than the period uncertainty alone suggests. Equilibrium temperature scales
as $T_{\rm eq} \propto P^{-1/3}$, so a period interval spanning a factor of
several compresses to a much narrower temperature interval. A period too
uncertain to schedule a follow-up observation may still be precise enough
to determine whether a planet could support liquid water.

This paper makes four claims, each tested against confirmed planets with
published periods:

\begin{enumerate}
\item Periodogram methods fail structurally on single transits, and the
      failure is exact rather than approximate.
\item The physics inverts a single transit with no period search grid.
\item Direct inversion underestimates systematically. Marginalising over
      the unobserved geometry removes the bias and halves the error.
\item Habitability assessment survives the residual period uncertainty
      because of the cube-root compression.
\end{enumerate}

Detection is not addressed. We assume the transit has been found and
localised, which is the problem \citet{priyanshu2026} addresses and
explicitly separates from period recovery.

\section{Methods}

\subsection{Data}

Light curves were retrieved from the Mikulski Archive for Space Telescopes
using \texttt{lightkurve} \citep{lightkurve2018}, preferring SPOC 120\,s
products. Stellar parameters came from the TESS Input Catalog
\citep{stassun2019} except where a discovery paper supersedes them.

For long-period planets most sectors contain no transit. Sectors were
selected by querying MAST observation metadata for each sector's time
coverage and comparing against the published ephemeris, choosing the
sector minimising $|n|$, the number of orbital cycles from the measured
epoch. Since propagated timing uncertainty
$\sigma_t(n) = \sqrt{\sigma_{T_0}^2 + (n\sigma_P)^2}$ grows monotonically
with $|n|$, this also minimises the risk of a mispredicted epoch. Where
the lowest-$|n|$ sector fell below the detection threshold, later sectors
were tried in ascending $|n|$.

This is a retrospective validation design. Selecting the sector requires
the published ephemeris, so the true period is used to construct the
evaluation set, and the results below measure period recovery given a
correctly localised transit rather than blind performance on unlabelled
data. This is unavoidable when validating against known answers. In
deployment the sector would be identified by detection significance alone,
and the measurement itself is unaffected: depth and duration are extracted
from the photometry without reference to the assumed period.

\subsection{Photometric extraction}

Extraction proceeds in two passes. The first detrends with a
Savitzky--Golay filter over a 5-day window with iterative $3\sigma$
clipping, locates the deepest coherent dip, and measures a provisional
duration. The second masks a window scaled to that duration, re-detrends,
and re-measures, with the mask width iterating to convergence.

\begin{figure}[H]
\centering
\includegraphics[width=0.9\textwidth]{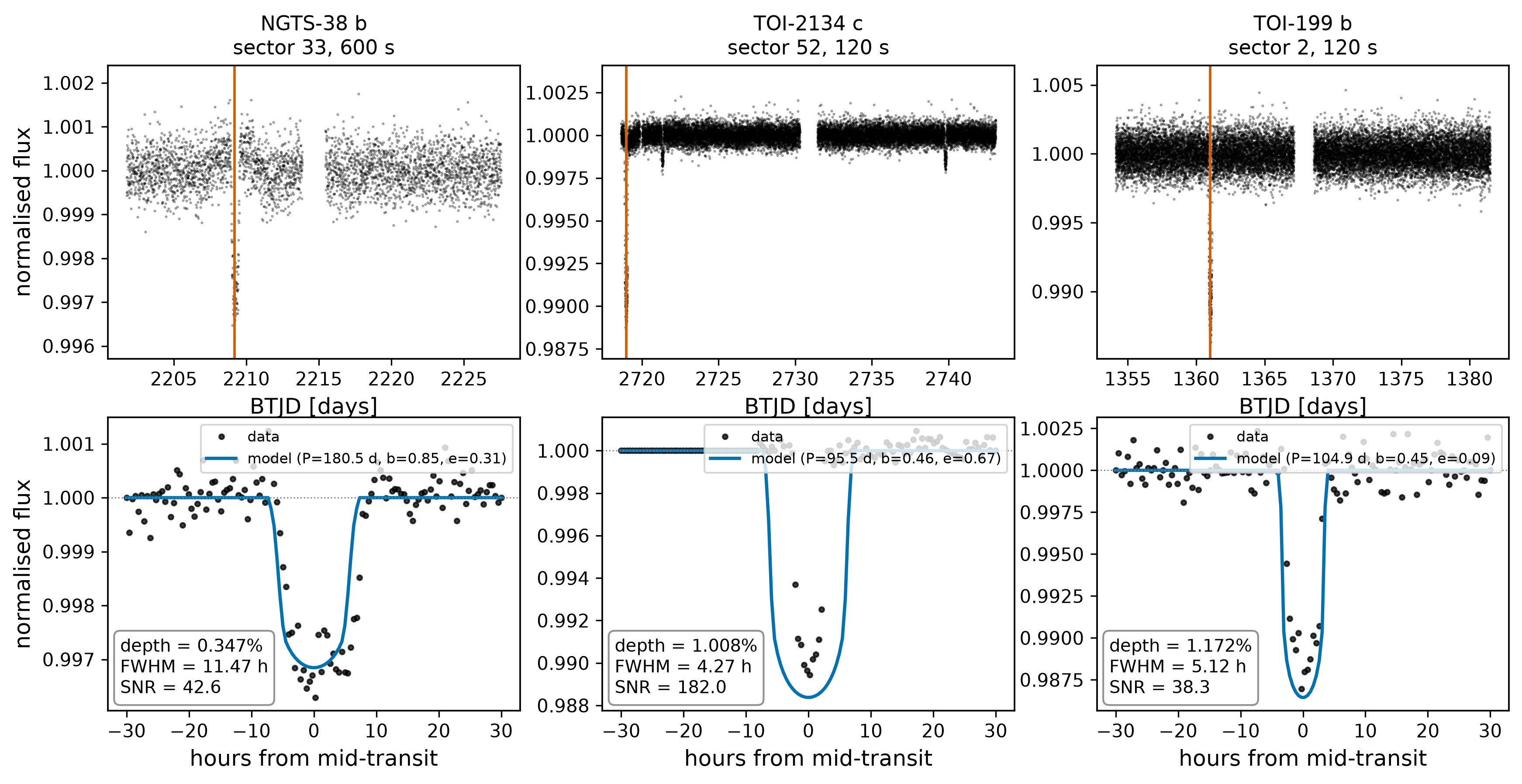}
\caption{Detrended light curves for the confirmed targets. Upper panels
show the full sector with the transit marked; lower panels show the
transit with the fitted model overlaid, annotated with measured depth and
FWHM.}
\label{fig:lightcurves}
\end{figure}

Extraction accuracy was validated by injecting transits of known depth and
duration into real photometry. Recovered durations agree with injected
values to within 1.4\% across 2--20\,h, and to 0.5\% at 15\,h.
Figure~\ref{fig:lightcurves} shows the extracted transits.

Targets were required to exceed an integrated signal-to-noise threshold of
$Q = 7.5$. This follows from duration precision rather than detection: the
FWHM is set by two half-depth crossings, at which a flux error maps to a
timing error, giving $\sigma_T/T = \sqrt{2}\sqrt{\tau/T}/Q$. For the
geometries in this sample $\tau/T \approx 0.09$, so
$\sigma_T/T \approx 0.45/Q$, and $Q = 7.5$ corresponds to 6\% duration
precision and hence 18\% period precision. Conventional detection requires
$Q \approx 7.1$.

\subsection{Forward model}

The total transit duration is
\begin{equation}
T_{14} = \frac{P}{\pi}\arcsin\!\left[\frac{R_\star}{a}
\sqrt{(1+k)^2 - b^2}\,\right],
\label{eq:t14}
\end{equation}
with $k = R_p/R_\star = \sqrt{\delta}$ from the transit depth $\delta$
\citep{winn2010,seager2003}, and $a$ from Kepler's third law,
\begin{equation}
a = \left(\frac{GM_\star P^2}{4\pi^2}\right)^{1/3}.
\label{eq:kepler}
\end{equation}

Inverting these for period gives $P \propto T^3$ in the small-angle limit,
so a duration ratio $r$ propagates to a period error of $r^3$. This
sensitivity governs everything that follows.

An eccentric orbit rescales the duration by
\begin{equation}
\frac{\sqrt{1-e^2}}{1 + e\sin\omega},
\label{eq:ecc}
\end{equation}
degenerate with the impact parameter dependence in
Equation~\ref{eq:t14}. Neither $b$ nor $e$ is recoverable from a single
transit at TESS photometric precision.

\subsection{Baselines}

\textit{Box Least Squares} was run using
\texttt{astropy.timeseries.BoxLeastSquares} with two grids: 0.5--27\,d,
the widest justifiable from a single sector, and 0.5--200\,d, deliberately
generous. Signal Detection Efficiency is reported throughout.

\textit{Direct inversion} solves Equations~\ref{eq:t14} and
\ref{eq:kepler} for $P$ by root finding, assuming $b = 0$ and $e = 0$. It
requires no training and no grid, and is the natural baseline against
which any learned method must justify itself.

\subsection{Network}

The model takes four scalars --- transit duration, transit depth, stellar
mass, stellar radius --- and outputs a mean and variance in $\log P$ under
a heteroscedastic Gaussian negative log-likelihood with a floor on the
predicted variance. A physics term penalises predictions inconsistent with
Equation~\ref{eq:t14}. It is applied as a tolerance band rather than an
equality because limb darkening leaves a $\sim$6\% residual between the
analytic duration and the measured FWHM: the penalty is zero while the
predicted duration lies within a fractional tolerance $\epsilon$ of the
measured value and grows quadratically outside it, weighted by $\lambda$
relative to the likelihood term. Values are given in
Appendix~\ref{app:config}.

Training draws $(P, M_\star, R_\star, k, b, e, \omega)$ from population
priors, evaluates Equation~\ref{eq:t14}, and measures depth and duration
with the same estimator used on real data. The result is a sample from the
joint distribution of observables and period, so a network trained under a
Gaussian NLL converges to the conditional posterior. The network is
amortised Bayesian inference: it performs in microseconds what explicit
Monte Carlo sampling performs in tens of seconds.

Impact parameter is drawn uniformly on $[0, 1+k]$, which is the correct
distribution for isotropic inclinations conditioned on a transit
occurring. Eccentricity follows a Beta distribution \citep{kipping2013}
with 25\% of systems forced circular, and $\omega$ is uniform. Periods are
floored at 3.0\,d; below this a second transit enters the extraction
window and the network learns to measure the interval between transits
rather than inverting the duration of one --- a different and considerably
easier problem (Section~\ref{sec:spacing}).

Full configuration is in Appendix~\ref{app:config}.

\subsection{Habitability}

Equilibrium temperature follows from the recovered period as
\begin{equation}
T_{\rm eq} = T_\star \sqrt{\frac{R_\star}{2a}}\,(1-A)^{1/4},
\end{equation}
with $A = 0.3$. Since $a \propto P^{2/3}$, temperature scales as
$P^{-1/3}$ and the period posterior maps to a temperature posterior
compressed by the cube root.

\subsection{Sample}

Of 30 confirmed single-transit planets selected, 3 had no observed sector
containing a predicted transit, 10 fell below the detection threshold in
the sector containing their transit, and 1 lies outside the model's
training prior, leaving $n = 16$. The 10 non-detections are a property of
the population rather than of the pipeline: a duration cannot be measured
from a transit that is not detected, and this bounds where any
single-transit method can operate.

\section{Results}

\subsection{Periodogram methods fail exactly, not approximately}
\label{sec:bls_fails}

\begin{figure}[H]
\centering
\includegraphics[width=0.75\textwidth]{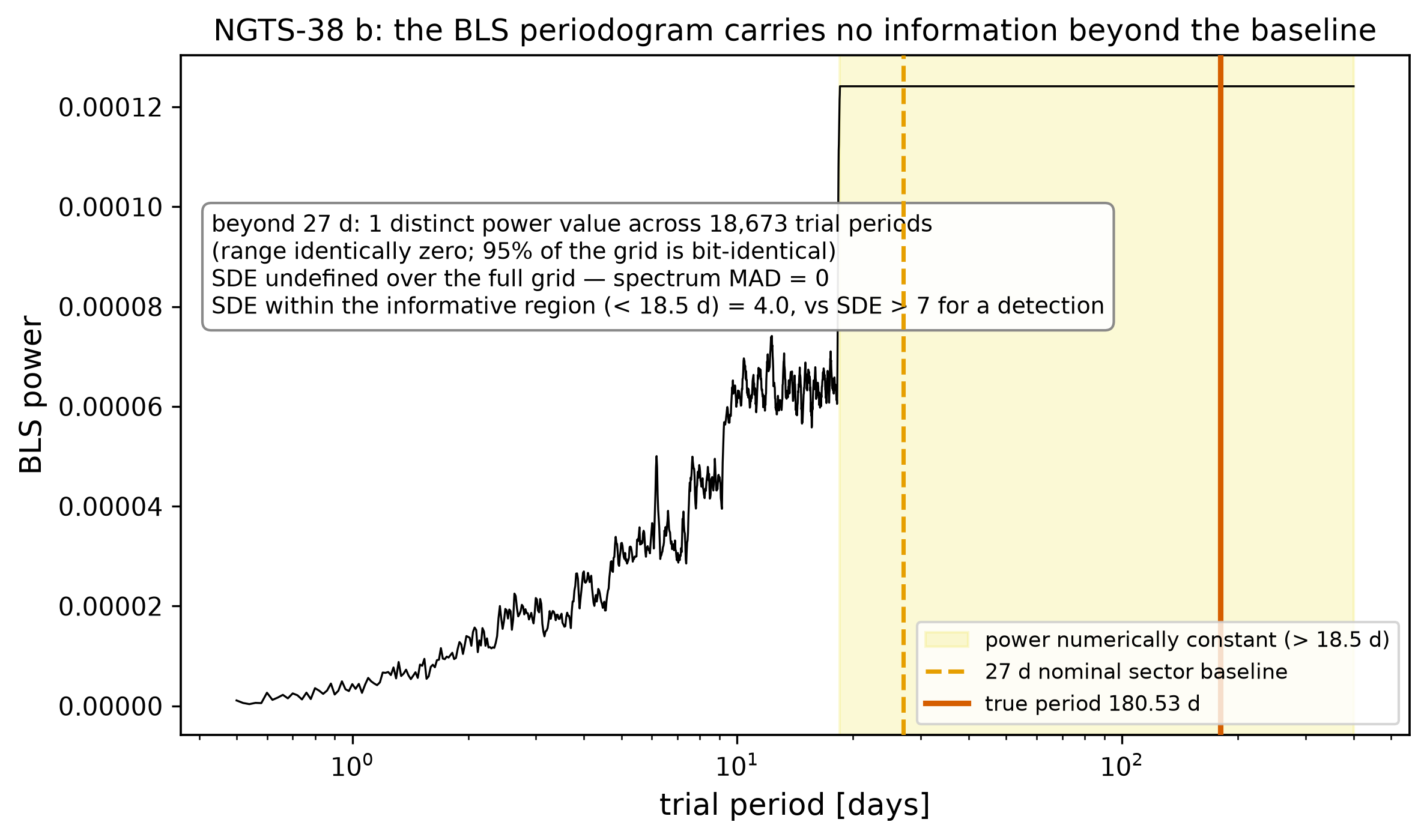}
\caption{BLS power spectrum for NGTS-38\,b. Beyond 18.5\,d --- the longest
contiguous span in the sector --- the power is numerically constant across
18{,}673 trial periods. The true period at 180.53\,d lies in this region.}
\label{fig:flat}
\end{figure}

Figure~\ref{fig:flat} shows the BLS power spectrum for NGTS-38\,b. Beyond
18.5\,d the power takes exactly one distinct value across 18{,}673 trial
periods. The spectrum's median absolute deviation is identically zero over
95\% of the grid, so the Signal Detection Efficiency is undefined and
\texttt{argmax} returns whichever grid point floating-point ordering
happens to favour.

The threshold at 18.5\,d is not the nominal 25.8\,d sector length. The
mid-sector downlink gap means the longest contiguous span is 18.5\,d, and
it is contiguous coverage that carries periodicity information. The
nominal baseline overstates BLS's reach by roughly 30\%.

\begin{figure}[H]
\centering
\includegraphics[width=0.75\textwidth]{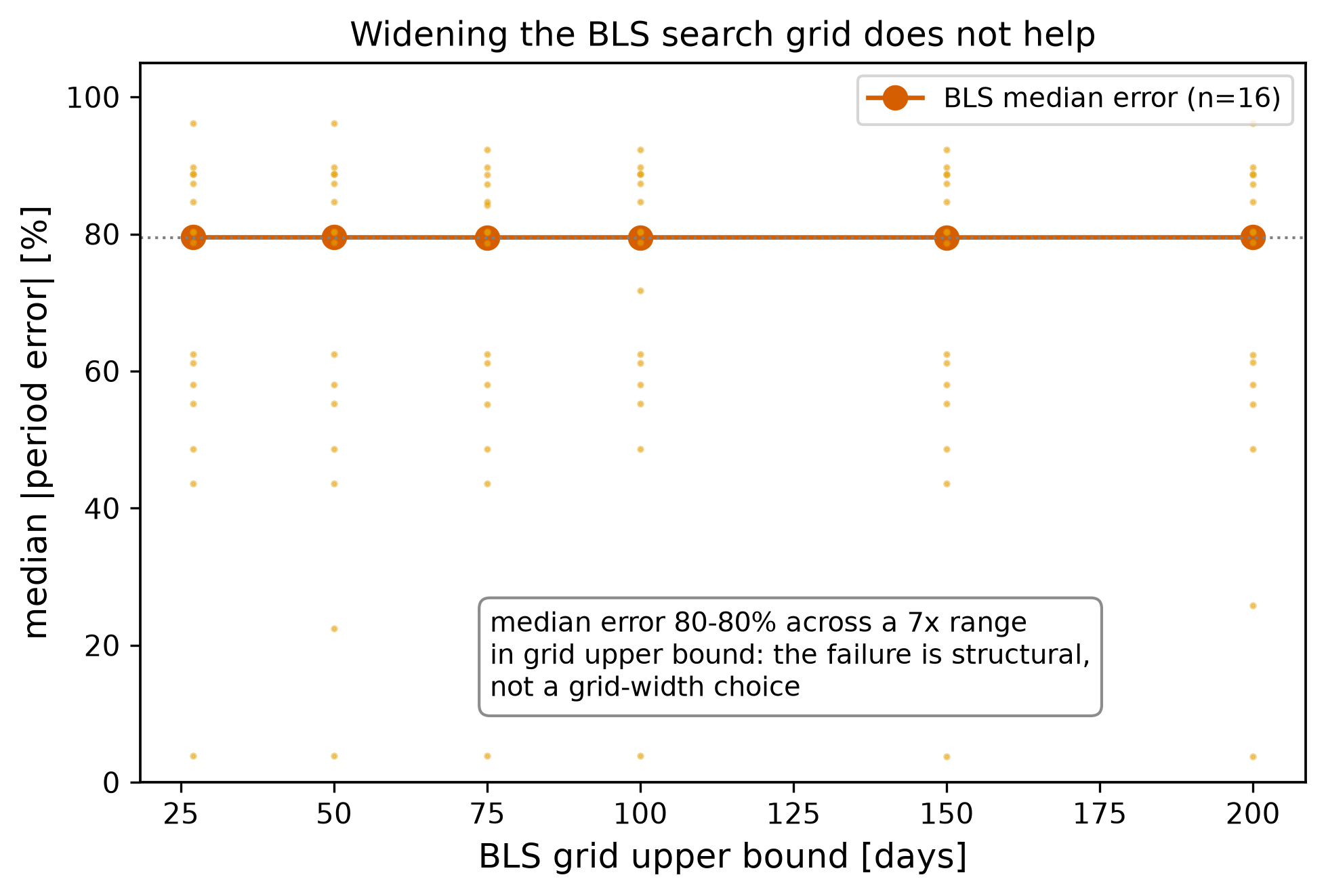}
\caption{Median BLS error across the 16 targets against the upper bound of
the search grid. The response is flat: the grid is irrelevant because the
periodogram carries no information.}
\label{fig:grid}
\end{figure}

Two consequences follow, both measured across the 16 targets. Widening the
search grid from 0.5--27\,d to 0.5--200\,d changes the median error by
$-0.0$ percentage points (Figure~\ref{fig:grid}); a method extracting
information would respond to a sevenfold increase in search space. And no
target yields a detection above SDE 4, against a conventional threshold of
7. BLS returned periods between 8.9 and 25.2\,d for every target,
irrespective of true periods spanning 26 to 232\,d.

No grid choice repairs this. BLS requires a second transit.

\subsection{Period recovery without a grid}

Table~\ref{tab:headline} gives median absolute error across the 16
targets.

\begin{table}[htbp]
\centering
\caption{Median absolute error, 16 confirmed single-transit planets.}
\label{tab:headline}
\begin{tabular}{lrr}
\toprule
Method & Median & Coverage \\
\midrule
BLS, 0.5--27\,d     & 79.5\% & --- \\
BLS, 0.5--200\,d    & 79.5\% & --- \\
Direct inversion    & 69.2\% & --- \\
Network             & 47.8\% & 14/16 \\
Network (injection) & \textbf{40.5\%} & 14/16 \\
\bottomrule
\end{tabular}
\end{table}

\begin{figure}[H]
\centering
\includegraphics[width=0.75\textwidth]{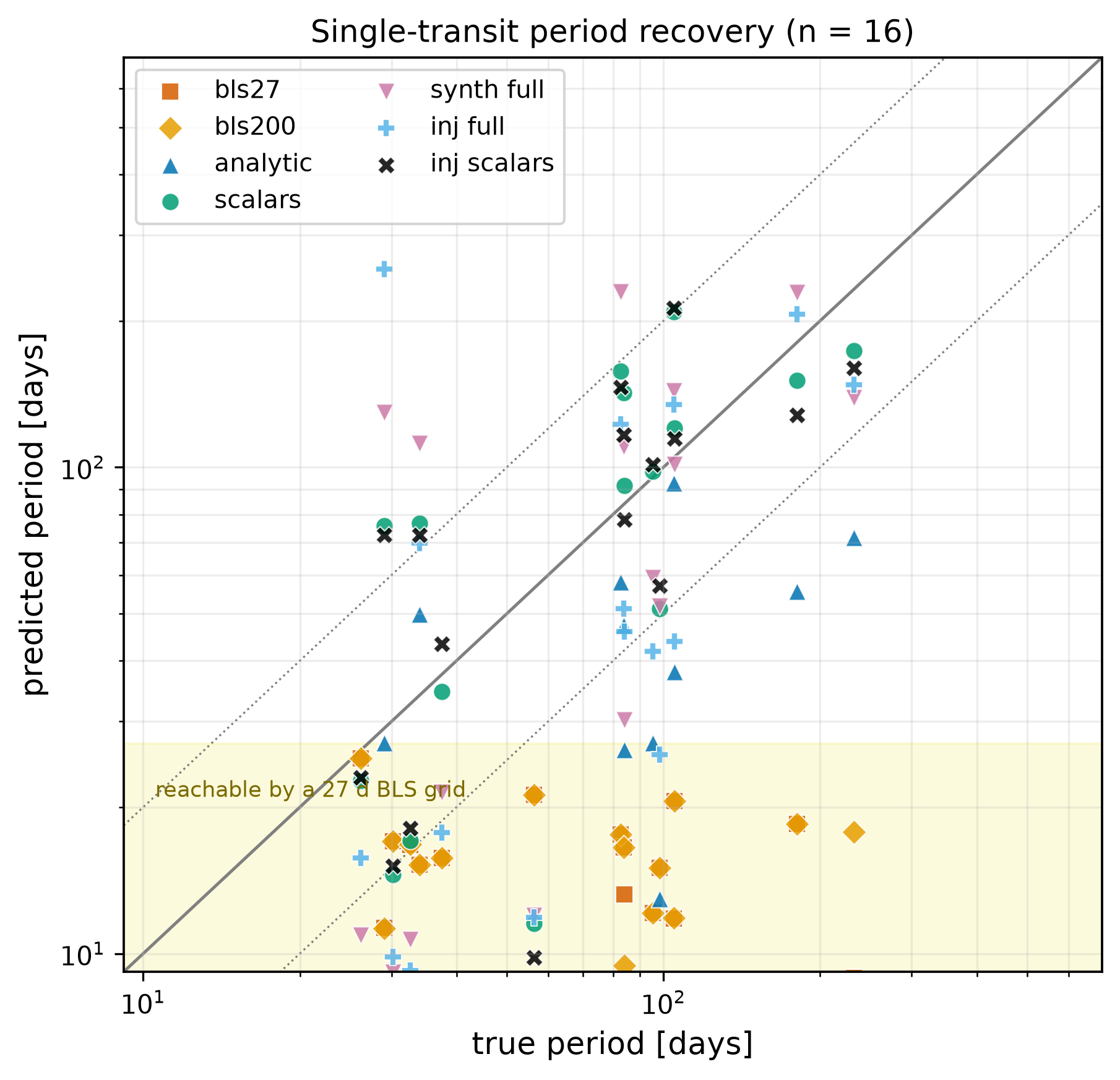}
\caption{Predicted against true orbital period for the 16 confirmed
targets, all methods. The diagonal marks perfect recovery. BLS predictions
cluster between 9 and 25\,d irrespective of the true period.}
\label{fig:predvstrue}
\end{figure}

The network halves the error against BLS and improves on direct inversion
by 29 percentage points, with the true period inside the $1\sigma$
interval for 14 of 16 targets.
Figure~\ref{fig:predvstrue} shows the same result per target.

\subsection{Why direct inversion fails, and what the network corrects}
\label{sec:mechanism}

Direct inversion underestimates the period in 14 of 15 targets with
published $b$, with a median signed error of $-69.3\%$. The network's
signed error is $-7.7\%$: essentially unbiased.

The bias follows from Equation~\ref{eq:t14}. Setting $b = 0$ maximises the
chord across the stellar disc, so for a fixed observed duration it returns
the smallest orbital separation and hence the shortest period consistent
with the data. Direct inversion can therefore only underestimate.

The network's correction is derivable from its prior. Since
$\log P = 3\log T - 3\log(\mathrm{chord}) + \mathrm{const}$, a model
trained under a Gaussian likelihood in log-space learns
$\mathbb{E}[\log(\mathrm{chord})]$ over the $b$ prior. For $b$ uniform on
$[0, 1+k]$,
\begin{equation}
\mathbb{E}\!\left[\log\sqrt{1-u^2}\right] = \ln 2 - 1 = -0.3069,
\end{equation}
giving a geometric-mean chord of $0.736\times$ the $b=0$ chord --- exactly
an effective impact parameter $b^\ast = 0.745$ at $k = 0.1$. Solving for
the $b^\ast$ reproducing each network prediction gives a median of 0.749,
agreeing with the derived value to 0.5\%. The value the true periods
require is 0.735.

Two things follow. The correction is principled rather than empirical: it
is the $b$-marginalised posterior the network was trained to represent,
and uniform $b$ is the correct prior for isotropic inclinations given a
transit. And it is population-level rather than per-target: the fitted
$b^\ast$ has an interquartile range of 0.693--0.772 and is uncorrelated
with published $b$ ($\rho = +0.007$, $p = 0.98$). The network does not
infer $b$; it applies the correction appropriate to a typical geometry.

\begin{figure}[H]
\centering
\includegraphics[width=0.75\textwidth]{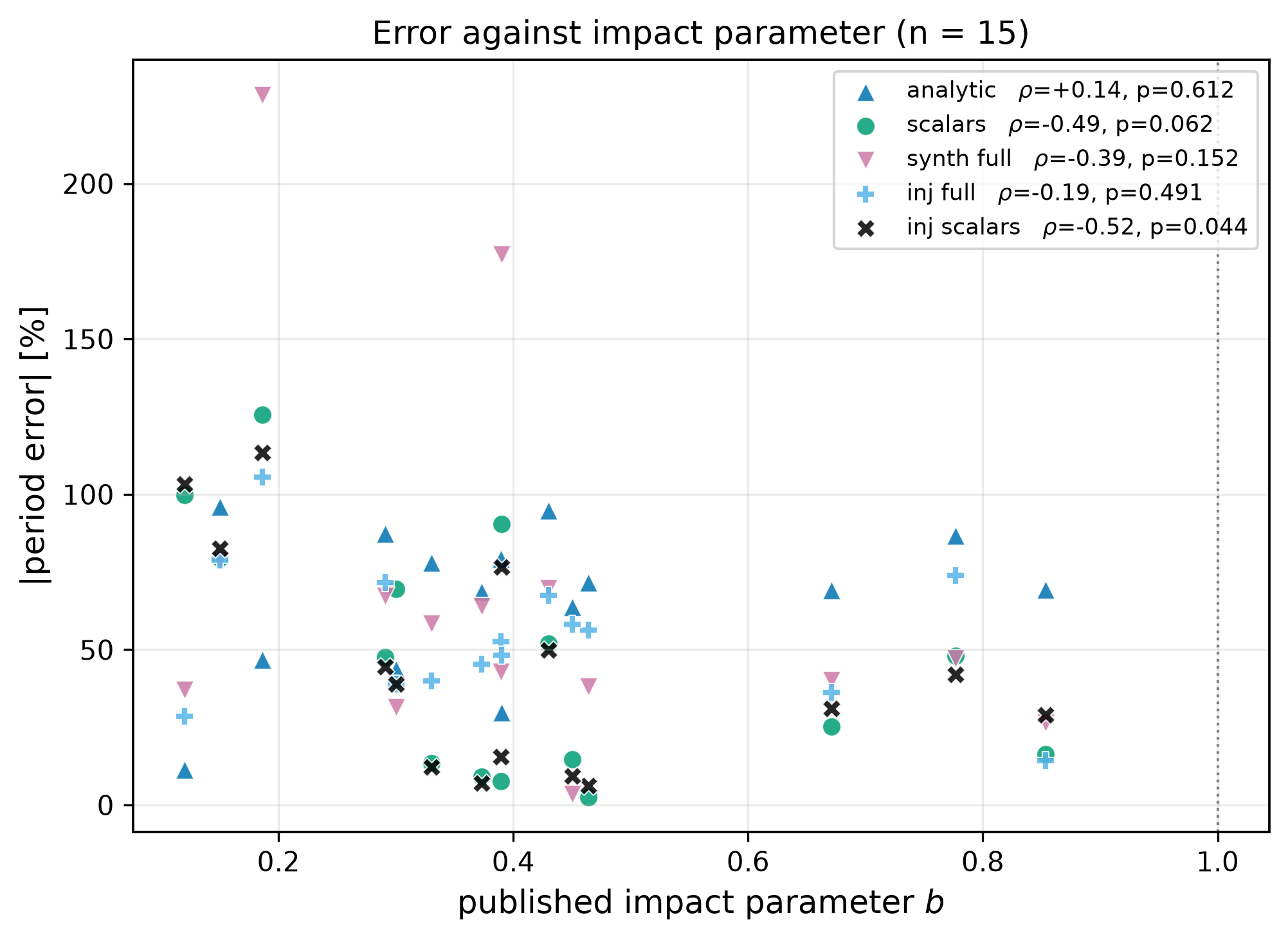}
\caption{Absolute error against published impact parameter for each
method, with Spearman rank correlations annotated.}
\label{fig:errvsb}
\end{figure}

\begin{figure}[H]
\centering
\includegraphics[width=0.75\textwidth]{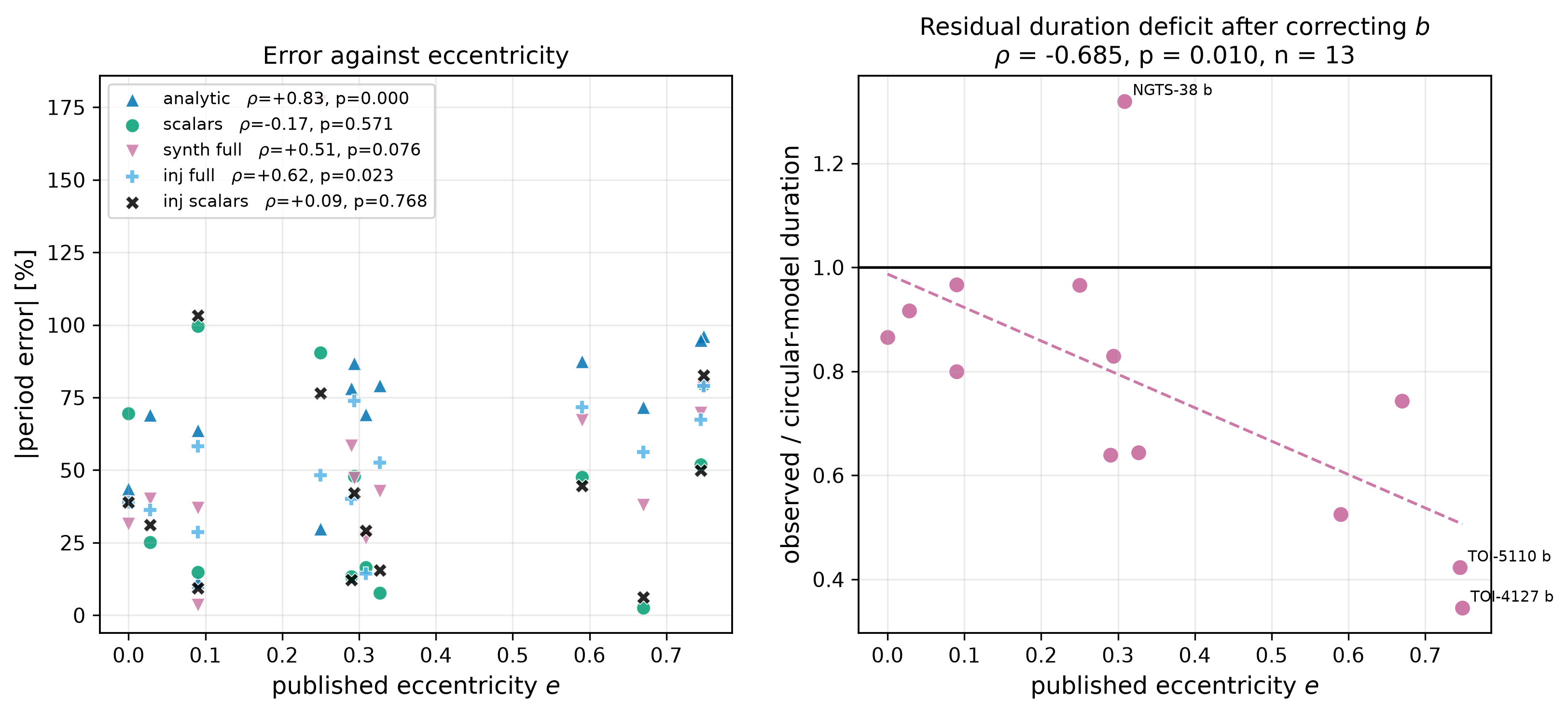}
\caption{Absolute error against published eccentricity. The residual
scatter grows with $e$, which is unobservable from a single transit.}
\label{fig:errvse}
\end{figure}

This explains why the bias vanishes while the scatter does not. Median
absolute error remains 40.5\% because eccentricity varies between targets,
shifts the duration through Equation~\ref{eq:ecc}, and is invisible from a
single transit. After correcting for true $b$, the residual duration
deficit correlates with published eccentricity at $\rho = -0.685$
($p = 0.010$, $n = 13$). The network gets the population median right and
cannot get individual targets right, which is the expected behaviour of an
amortised posterior over unobservable parameters. Figures~\ref{fig:errvsb} and~\ref{fig:errvse} show both relationships.

\begin{figure}[H]
\centering
\includegraphics[width=0.75\textwidth]{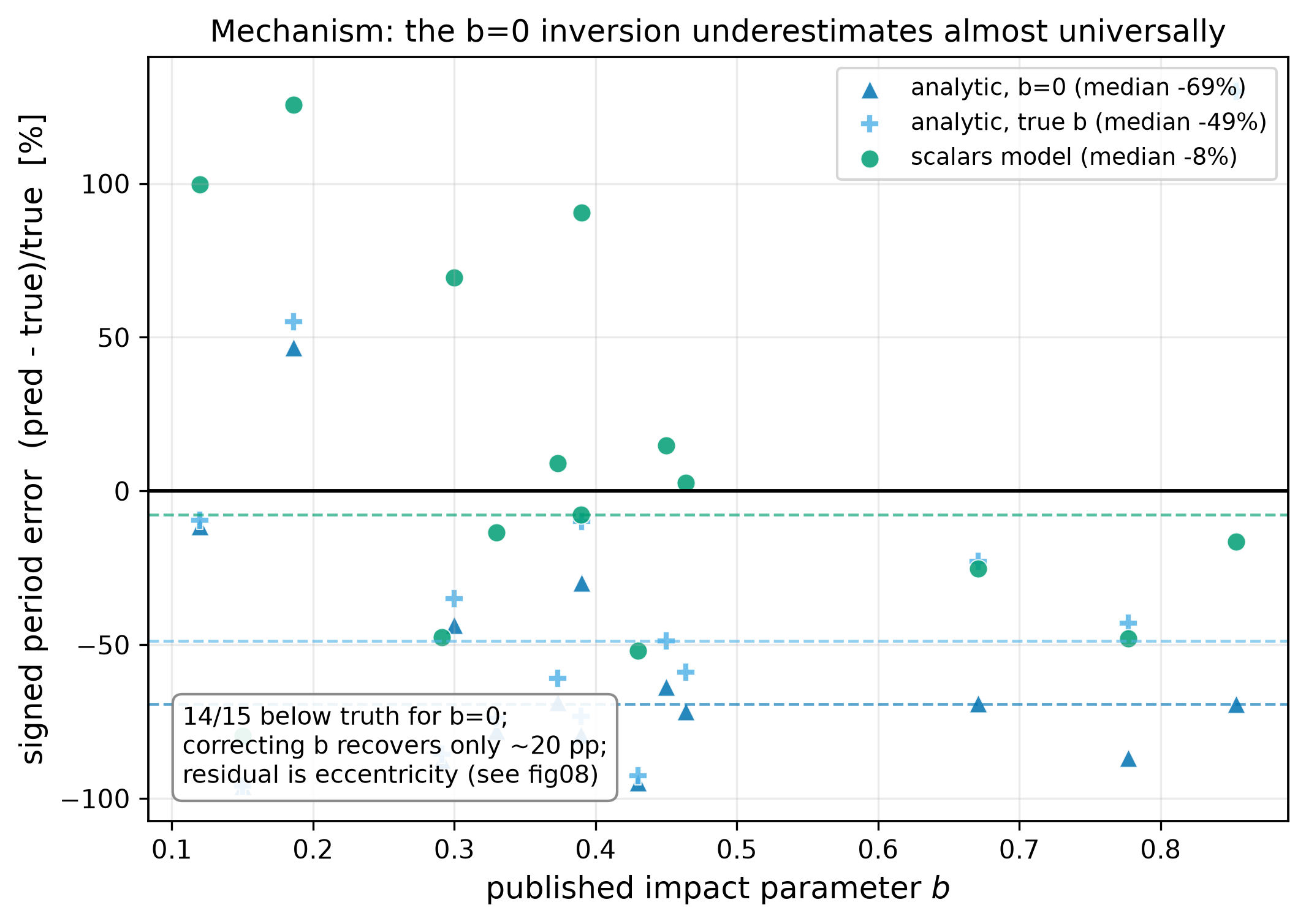}
\caption{Signed error against published impact parameter. Direct inversion
underestimates for all but one target across the full range of $b$
(median $-69\%$); the network's errors scatter in both directions with a
median of $-7.7\%$.}
\label{fig:mechanism}
\end{figure}

\subsection{NGTS-38\,b: temperature survives the period uncertainty}

NGTS-38\,b has a published period of 180.53\,d, an order of magnitude
beyond any grid a single sector could justify. Its transit appears in
sector 33 as a 0.335\% dip at $7.9\times$ the local scatter, against a
published depth of 0.349\%. Sectors 7, 61, 87 and 88 show no signal above
noise.

The measured FWHM is 11.52\,h. A circular orbit predicts 8.69\,h; folding
in the published $e = 0.309$ with transit near apoastron predicts
11.95\,h, a ratio of 0.96. The measurement and the eccentric forward model
agree.

Table~\ref{tab:degeneracy} shows the consequence of the unobserved
geometry. The same measured duration is consistent with periods from 39 to
490\,d, a factor of 12.5, and only the true geometry recovers the true
period.

\begin{table}[htbp]
\centering
\caption{Period implied by NGTS-38\,b's measured duration under different
geometric assumptions.}
\label{tab:degeneracy}
\begin{tabular}{lrr}
\toprule
Assumption & Implied $P$ & vs truth \\
\midrule
$b=0$, $e=0$ & 101.5\,d & 0.56$\times$ \\
$b=0.854$, $e=0$ & 489.5\,d & 2.71$\times$ \\
$b=0$, $e=0.309$ & 39.0\,d & 0.22$\times$ \\
Published $b$ and $e$ & 188.0\,d & 1.04$\times$ \\
\bottomrule
\end{tabular}
\end{table}

Marginalising over $b$, $e$ and $\omega$ with 20{,}000 draws and no period
grid gives a posterior median of 190.7\,d against a truth of 180.53\,d,
with the true value at the 47.8th percentile.

\begin{figure}[H]
\centering
\includegraphics[width=0.75\textwidth]{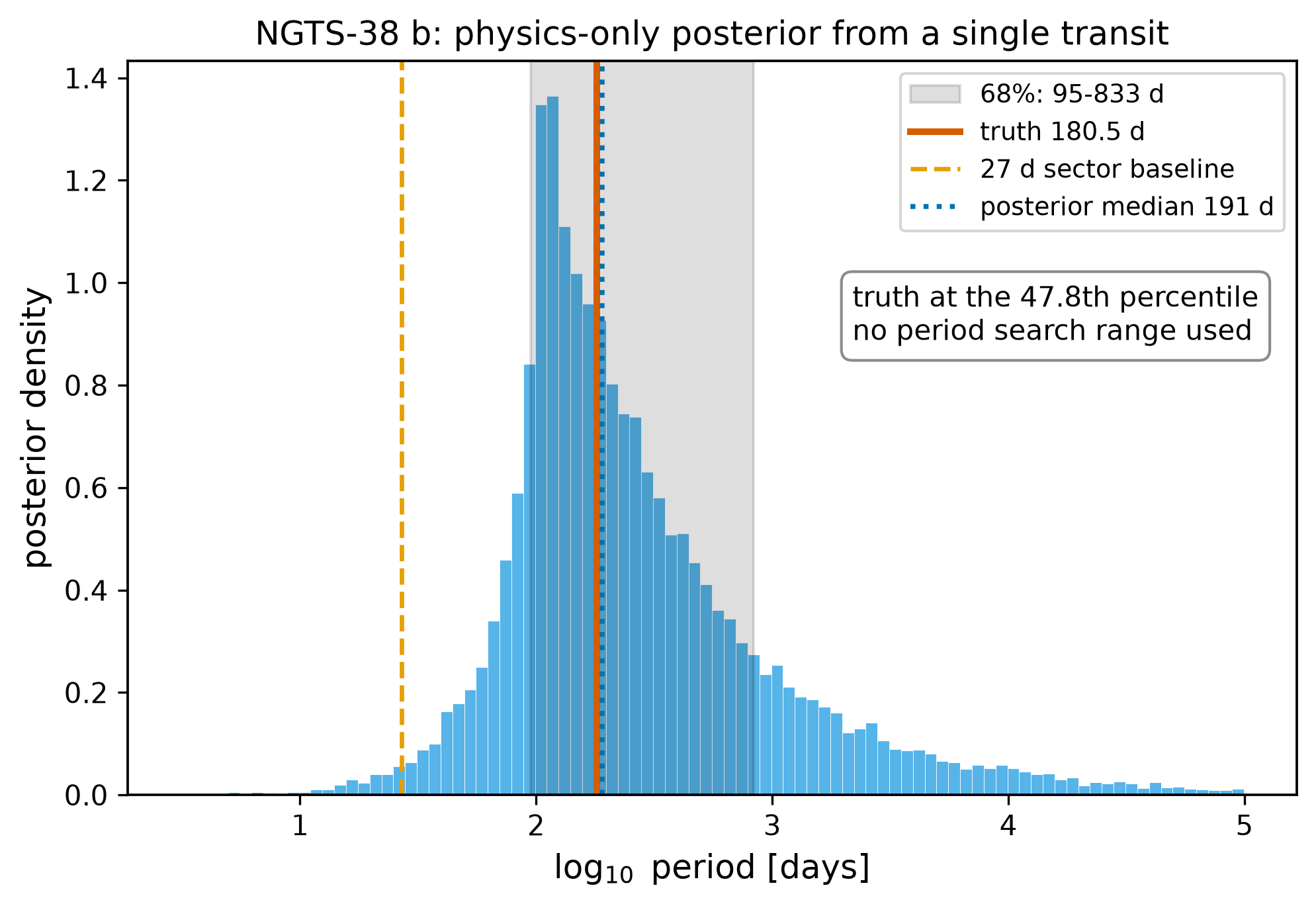}
\caption{Period posterior for NGTS-38\,b from marginalising over $b$, $e$
and $\omega$ with no search grid. The true period at 180.53\,d falls at
the 47.8th percentile.}
\label{fig:posterior}
\end{figure}

The corresponding equilibrium temperature posterior has a median of 445\,K
with a 68\% interval of 272--562\,K. The period interval spans a factor of
8.8; the temperature interval spans 2.1. This is the practical consequence
of the cube-root scaling: a period too uncertain to schedule follow-up
photometry still constrains temperature to a factor of two, and the lower
bound of the interval reaches into the conventional habitable zone, so a
habitable-zone orbit cannot be excluded at $1\sigma$.

\begin{figure}[H]
\centering
\includegraphics[width=0.75\textwidth]{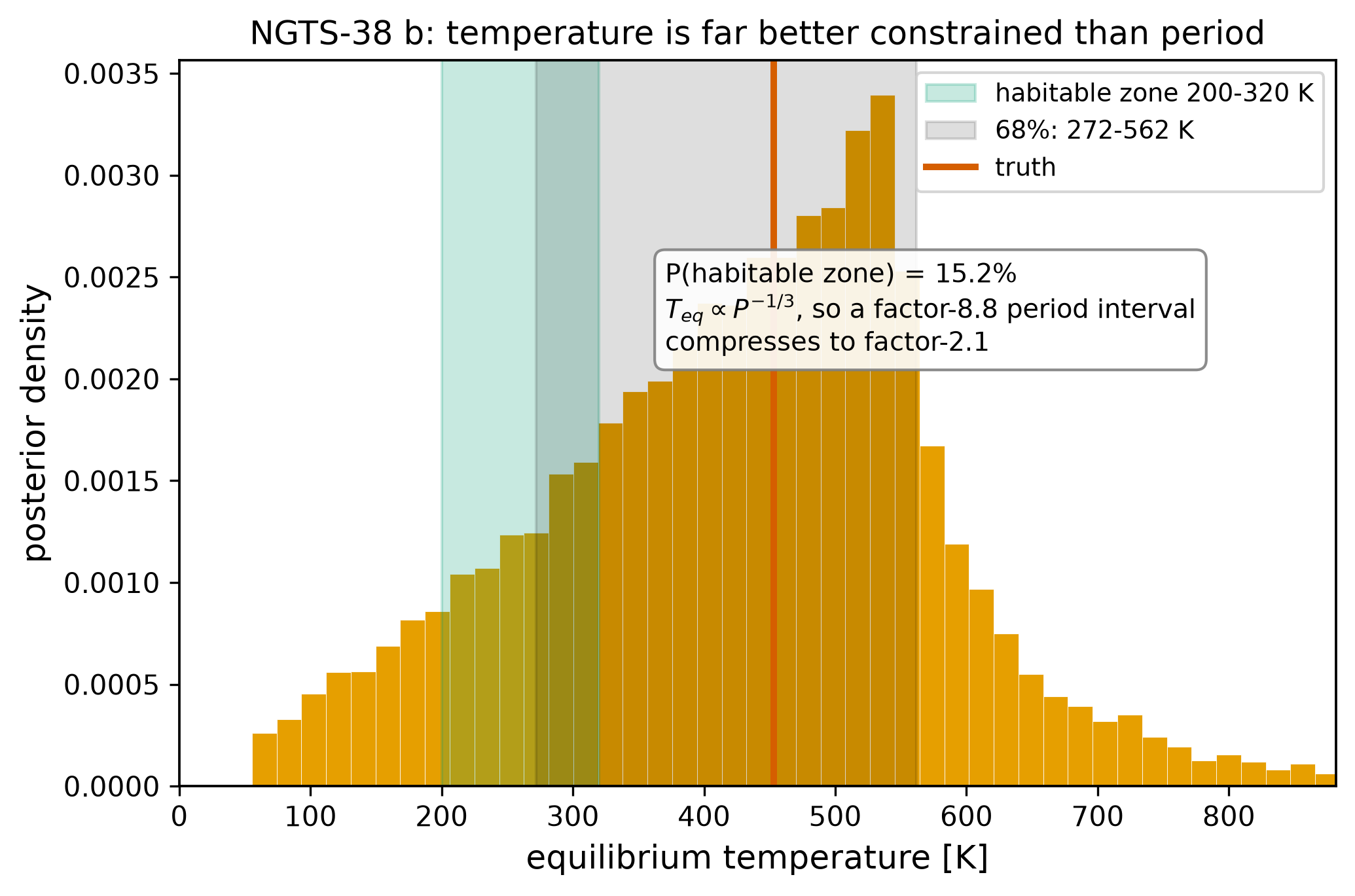}
\caption{Equilibrium temperature posterior for NGTS-38\,b from a single
transit. Shading marks the conventional habitable zone.}
\label{fig:temp}
\end{figure}

\subsection{Unconfirmed candidates}

The five single-transit candidates in Table~\ref{tab:cand} have no
published period, so no error can be computed. They are included to show
what the method produces on the case it is intended for: a real transit
whose period is genuinely unknown. All five clear the detection threshold.

\begin{table}[htbp]
\centering
\caption{Measurements and period predictions for five unconfirmed
single-transit candidates. No published period exists for any of them.}
\label{tab:cand}
\begin{tabular}{lrrrrr}
\toprule
TIC & Depth & FWHM & SNR & BLS & Network \\
    & (\%)  & (h)  &     & (d) & (d) \\
\midrule
233577004 & 0.552 & 8.71  & 42.4 & 21.1 & 97.8 \\
341687821 & 0.167 & 9.48  & 26.1 & 19.3 & 158.3 \\
122522333 & 0.201 & 7.51  & 24.0 & 16.8 & 58.6 \\
438122862 & 0.161 & 15.40 & 16.1 & 14.3 & 242.9 \\
232616346 & 0.308 & 11.55 & 35.8 & 14.9 & 106.8 \\
\bottomrule
\end{tabular}
\end{table}

Box Least Squares returns 14.3--21.1\,d for all five, with signal
detection efficiencies of 0.7--1.9 against a conventional threshold of 7.
The durations span a factor of two, from 7.5 to 15.4\,h, and the relation
between the two is inverted: the longest transit receives the shortest
period. Since $P \propto T^3$, the physical relation runs the other way.
This is the behaviour described in Section~\ref{sec:bls_fails} --- the
returned values reflect the search grid rather than the data.

\begin{figure}[H]
\centering
\includegraphics[width=0.9\textwidth]{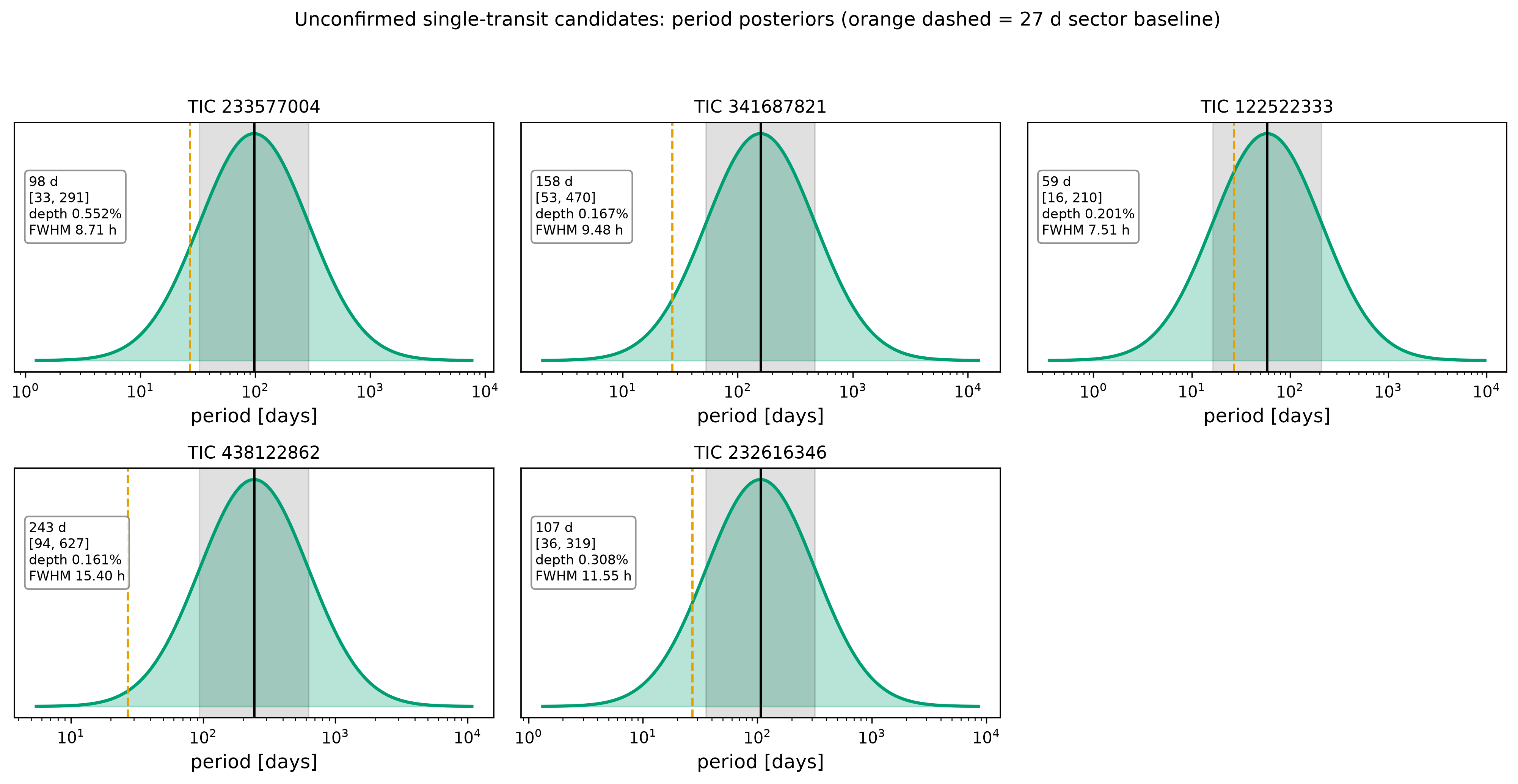}
\caption{Period posteriors for the five unconfirmed candidates. No
published period exists for comparison.}
\label{fig:candidates}
\end{figure}

The network returns 58.6--242.9\,d, with $1\sigma$ intervals spanning
factors of six to nine, consistent with the eccentricity-driven scatter
established in Section~\ref{sec:mechanism}. Point estimates should not be
quoted without them. Direct inversion returns 15.5--120.4\,d,
systematically shorter, as expected from its $b = 0$ assumption.
Figure~\ref{fig:candidates} shows the full posteriors.

\begin{table}[htbp]
\centering
\caption{Derived quantities at the network period. The habitable zone is
taken as $200 \le T_{\rm eq} \le 320$\,K.}
\label{tab:cand_temp}
\begin{tabular}{lrrrl}
\toprule
TIC & $a$ (AU) & $T_{\rm eq}$ & 68\% & HZ \\
\midrule
233577004 & 0.421 & 479 & 333--690 & no \\
341687821 & 0.607 & 408 & 284--587 & not excluded \\
122522333 & 0.314 & 651 & 426--997 & no \\
438122862 & 0.900 & 483 & 352--663 & no \\
232616346 & 0.433 & 502 & 349--724 & no \\
\bottomrule
\end{tabular}
\end{table}

No candidate has a central temperature estimate inside the habitable zone.
One, TIC~341687821, has a $1\sigma$ interval reaching 284\,K at its cold
edge, so a habitable-zone orbit cannot be excluded for it at the
long-period end of its posterior. The remaining four are excluded at
$1\sigma$. This is the intended output: not a period precise enough to
schedule a follow-up transit, but a temperature constraint sufficient to
decide which candidates merit one and which do not.

These five also illustrate the extraction sensitivity discussed in
Section~\ref{sec:extraction}. TIC~438122862's measured duration moved from
1.95\,h to 15.40\,h across the pipeline corrections listed in
Appendix~\ref{app:bugs}, a factor of eight. Propagated through
$P \propto T^3$, its direct-inversion period moved from 9.7 to 120.4\,d.
No difference between the methods compared in this paper approaches that
magnitude.

\section{Discussion}

\subsection{Transit shape does not transfer}
\label{sec:spacing}

A variant of the network additionally ingests the 128-point flux array
through a convolutional branch. On held-out synthetic data it improves
median error from 57.3\% to 38.2\%, concentrated where the physics
predicts: 46\% for circular orbits, 18\% for $0.15 < e < 0.35$, and none
above $e = 0.35$. Transit shape constrains impact parameter, and as
eccentricity comes to dominate, that constraint loses value.

The improvement does not survive contact with real photometry. On the 16
targets the convolutional model reaches 54.4\% against its scalars-only
control's 40.5\%, with interval coverage degrading from 14/16 to 8/16.
Injecting synthetic transits into real light curves before detrending, so
that injected and real signals are distorted identically, did not close
the gap.

\citet{iglesias2024} report a convolutional network recovering periods
from TESS light curves at 1.87\% error on real data, trained on synthetic
data cruder than ours. Their Table~1 gives a training period range of
1--13.5\,d, and their Section~2.2 states that at least two transits were
injected per light curve. Their real test set spans 0.94--7.86\,d. Every
example contains multiple transits, and their network measures the
interval between them.

That distinction is the point. Transit spacing is a time difference
between high signal-to-noise events, insensitive to photometric
distortion. Transit shape is not: the ingress slope and flat-bottom
curvature encoding $b$ sit near the noise floor for depths below
$\sim$1\%, and detrending distorts them directly. Single-transit period
recovery has no spacing available and must rely on shape --- precisely the
observable that fails to transfer.

We encountered the same effect from the other direction. An early training
set used a 2.5\,d period floor, admitting a second transit into the
extraction window for $\sim$13\% of samples. Performance improved, because
the network had learned to read period from spacing. We raised the floor
to 3.0\,d, since it solves a different problem from the one posed here.

\subsection{Measurement dominates the error budget}
\label{sec:extraction}

The largest single effect in this work was not a change of model but a
change of measurement. Thirteen pipeline defects were identified and
corrected (Appendix~\ref{app:bugs}); several produced errors exceeding any
difference between methods. Selecting a sector by transit centrality
rather than by cycle count changed one target's measured depth by a factor
of 19, and hence its inferred period by more than two orders of magnitude.
For anyone attempting single-transit period recovery, photometric
extraction deserves more scrutiny than model selection.

\subsection{Limitations}

\textit{Sample size.} Statistical claims rest on the held-out synthetic
set. The 16-target result demonstrates that the pipeline recovers known
answers end-to-end; it is not a population statistic.

\textit{Selection.} The network's training prior contains no
transit-probability weighting, so the periastron bias present in a
transit-selected real sample is absent from training. The $b$-marginalised
correction happens to approximate what that bias requires, and this is not
guaranteed to hold for a differently selected sample.

\textit{Prior saturation.} The model cannot predict outside its 3--1000\,d
training prior. TOI-201\,c, with a published period of 2890\,d, was
excluded on this basis; the network returns 250.8\,d and the direct
inversion 135.3\,d, both saturating well below the truth. Any target
beyond the prior will be underestimated by construction, and the prior
sets the method's applicable range.

\textit{Stellar parameters.} Since $P \propto M_\star/R_\star^3$, stellar
parameter errors propagate directly. For NGTS-38\,b the discovery paper
gives $M_\star = 1.46\,M_\odot$ against the TIC's 1.191, moving the
inferred period from 143.6 to 188.0\,d --- a 20\% systematic beneath every
method compared here. We use the discovery paper value throughout.

\textit{Eccentricity.} The dominant residual uncertainty, marginalised
over in the prior but unconstrained by the data.

\subsection{Future work}

Three approaches address the shape-transfer failure. \citet{osborn2020}
generate synthetic TESS data at pixel level through the full SPOC
pipeline. \citet{fiscale2021} pretrain on Kepler, whose precision supports
learning shape features, and fine-tune on TESS. \citet{cuellar2022} train
on a mixture of real and synthetic data. Constraining eccentricity
requires a second observable --- spectroscopy, or a second transit, which
returns the problem to where BLS already works.

\section{Conclusion}

Single-transit planets cannot have their periods measured by periodogram
methods, and the failure is exact: the power spectrum is numerically
constant across the region containing the answer. Recovering the period
from transit duration through Kepler's third law requires no search grid,
but a direct inversion underestimates by a median of 69\% because assuming
a central transit returns the shortest period consistent with the data. A
network marginalising over the unobserved geometry removes that bias and
reaches 40.5\% median error against 79.5\% for Box Least Squares.

The residual uncertainty is set by eccentricity, which is invisible from a
single transit. But because equilibrium temperature scales as $P^{-1/3}$,
that uncertainty compresses: for NGTS-38\,b, a period known only to within
a factor of 8.8 yields a temperature known to within a factor of 2.1,
enough to place the planet relative to the habitable zone from one
observation.

\section*{Use of AI tools}
The software developed for this work and the drafting of this manuscript
were carried out with the assistance of Anthropic's Claude. The research
questions, experimental design, and all scientific judgements are the
author's own, as is responsibility for the correctness of the results
presented here. All code is available in the repository linked below and
all numerical results are reproducible from it.

\section*{Data availability}
This paper includes data collected by the TESS mission, publicly available
from the Mikulski Archive for Space Telescopes. Planetary and stellar
parameters were obtained from the NASA Exoplanet Archive and the TESS
Input Catalog. All code used in this work --- photometric extraction, the
forward model, training, and the analysis reproducing every figure and
table --- is available at
\url{https://github.com/hassanxj21/single-transit-period-recovery},
together with the trained model weights and the target lists.

\section*{Acknowledgements}
This paper includes data collected by the TESS mission, funded by the NASA
Explorer Program. This research has made use of the NASA Exoplanet
Archive, operated by the California Institute of Technology under contract
with NASA under the Exoplanet Exploration Program, and of the Mikulski
Archive for Space Telescopes.

\bibliographystyle{plainnat}
\bibliography{refs}

\appendix

\section{Model and training configuration}
\label{app:config}

\subsection*{Training data}

The synthetic set contains 60{,}000 samples and the injection-recovery set
25{,}000, split 85/15 into training and validation. Priors are identical in
both generators:

\begin{table}[H]
\centering
\begin{tabular}{lll}
\toprule
Quantity & Form & Bounds \\
\midrule
$P$ & log-uniform & 3.0--1000\,d \\
$M_\star$ & uniform & 0.15--2.5\,$M_\odot$ \\
$R_\star$ & MS relation$^a$ & 0.10--6.0\,$R_\odot$ \\
$k$ & log-uniform & 0.02--0.30 \\
$b$ & $U(0,1.05)\times(1+k)$ & $0$--$(1+k)$ \\
$e$ & Beta(0.867, 3.03)$^b$ & 0--0.85 \\
$\omega$ & uniform & 0--$2\pi$ \\
$u_1, u_2$ & uniform & 0.20--0.60, 0.05--0.40 \\
$\sigma_{\rm noise}$ & log-uniform & $10^{-4}$--$3\times10^{-3}$ \\
\bottomrule
\end{tabular}
\end{table}

\noindent
$^a$ $R = M^{0.80}$ for $M \le 1$, $M^{0.57}$ above, with lognormal scatter
of 0.045\,dex; 15\% of systems are inflated by $U(1.6, 3.5)$ to represent
evolved stars.
$^b$ \citet{kipping2013}, clipped at 0.85, with 25\% of systems forced
exactly circular.

No transit-probability weighting is applied: every drawn system passing the
geometric and detectability cuts is retained, regardless of how likely that
$(e, \omega)$ combination would be to transit.

The extraction window is 2.5\,d sampled at 128 points, giving a cadence of
28.125\,min. The 3.0\,d period floor follows from this: any shorter period
places a second transit in the same array. Training samples require an
integrated signal-to-noise of at least 10, defined as
$(\delta/\sigma)\sqrt{N_{\rm in}}$ with $\sigma$ the scaled median absolute
deviation of the window and $N_{\rm in}$ the number of in-transit cadences,
and a duration no greater than 42\,h.

\subsection*{Architecture}

The scalars-only model has 30{,}338 trainable parameters. It takes four
standardised inputs --- $\log_{10}$ duration in hours, $\log_{10}$ depth,
$M_\star$, and $\log_{10} R_\star$ --- through fully connected layers of
width 64, 64, 128 (with dropout 0.10) and 64, all ReLU, to a two-element
output giving $\mu$ and $\log\sigma$.

The convolutional variant has 161{,}410 parameters. Its CNN branch applies
three Conv1d layers with 16, 32 and 64 channels and kernel sizes 7, 5 and
3, each with batch normalisation, ReLU and max-pooling, reducing the
128-point input to a 1024-element vector. This is concatenated with a
64-element scalar branch and passed through a head of width 128 (dropout
0.10) and 64.

\subsection*{Loss}

\begin{equation}
\mathcal{L} = \mathcal{L}_{\rm data} + \lambda\,\mathcal{L}_{\rm physics}
\end{equation}

with a heteroscedastic Gaussian likelihood
\begin{equation}
\mathcal{L}_{\rm data} = \left\langle \tfrac{1}{2}e^{-2\log\sigma}
(z - \mu)^2 + \log\sigma \right\rangle,
\end{equation}
where $z$ is standardised $\log_{10}P$, and a physics term
\begin{equation}
\mathcal{L}_{\rm physics} = \left\langle
\mathrm{ReLU}(|r| - \epsilon)^2 \right\rangle
+ 10\left\langle \mathrm{ReLU}(1 - a/R_\star)^2 \right\rangle,
\end{equation}
where $r = \ln T_{\rm model} - \ln T_{\rm obs}$ and $T_{\rm model}$ is the
circular central-transit duration at the predicted period. The second term
penalises orbits inside the star.

The tolerance is $\epsilon = 0.35$ in natural log, i.e.
$+41.9\%/-29.5\%$ fractional, with $\lambda = 0.1$ and a five-epoch
warm-up during which $\lambda = 0$. The variance floor is
$\log\sigma \ge -2.5$, equivalent to $\sigma \ge 0.059$\,dex.

\subsection*{Optimisation}

AdamW with weight decay $10^{-4}$, learning rate $1.5\times10^{-3}$ under a
cosine schedule to $3\times10^{-5}$, batch size 512, gradient clipping at
max-norm 5, up to 60 epochs with early stopping on validation NLL at
patience 12. The best-validation checkpoint is retained.

\begin{table}[htbp]
\centering
\begin{tabular}{lrrr}
\toprule
Model & Best ep. & Val NLL & Wall clock \\
\midrule
Full, synthetic     & 12 & $-0.355$ & 320\,s \\
Scalars, synthetic  & 23 & $-0.157$ & 8.6\,s \\
Full, injection     & 18 & $-0.244$ & 165\,s \\
Scalars, injection  & 51 & $+0.024$ & 7.4\,s \\
\bottomrule
\end{tabular}
\end{table}

All training ran on an Apple M1 CPU with four threads; no GPU was used.

\subsection*{Injection training}

The noise pool comprises 80 light curves of 40 distinct field stars, 2031
days of real photometry across 658{,}253 cadences, spanning $T_{\rm mag}$
8.00--10.36 (median 9.05) and 16 distinct sectors. Cadences are 120\,s (31
light curves), 200\,s (11) and 1800\,s (38). No evaluation target appears
in the pool, asserted in code.

Of 118{,}010 draws, 25{,}000 were retained (21.2\% acceptance). Transits
were injected into raw flux before detrending, into contiguous segments of
at least 8\,d, no closer than 1\,d to a segment edge and 3\,d to any
masked feature. Only the deepest dip in each pool light curve was masked;
no variability screen was applied.

\subsection*{Software}

Python 3.11.4, torch 2.13.0, numpy 2.4.6, scipy 1.17.1, astropy 8.0.1,
\texttt{lightkurve} 2.6.0, \texttt{astroquery} 0.4.11. Stellar parameters
are from TIC-8 as served by MAST \citep{stassun2019}. The NASA Exoplanet
Archive was queried on 2026-07-26 for target selection and 2026-07-27 for
ephemerides and references.

\section{Pipeline defects}
\label{app:bugs}

Thirteen distinct defects were identified and corrected during
development. Each changed published numbers, and several produced errors
exceeding any difference between the methods compared in this paper. They
are listed here because anyone implementing the same pipeline is likely to
encounter them.

\begin{enumerate}\itemsep2pt

\item \texttt{lightkurve}'s \texttt{flatten()} treats \texttt{mask=True}
as \emph{exclude from fit}, but the code passed the inverted mask, fitting
the trend through the transit and excluding the baseline. TOI-1899's depth
read 0.926\% against a published 4.45\%.

\item The transit mask spanned $\pm1.5$\,d while the filter window was
1\,d, so near the transit almost no unmasked data remained and the
interpolated trend sagged into the dip. Depths and durations came out
systematically short.

\item The first detrend pass used a 1-day window, shorter than long
transits, so the filter fitted the trend through the event. NGTS-38\,b's
11.4\,h transit measured 3.05\,h at 0.075\% depth.

\item The detrender omitted iterative sigma-clipping, which incidentally
protects transits by dropping points below the trend. NGTS-38\,b's depth
read 0.118\% against a published 0.349\%.

\item The filter window was set at $8\times$ the measured duration with no
cap, so a 40\,h transit produced a 14-day window whose edge effects
reached past the centre of a short segment but not of a full sector,
making the distortion depend on segment length.

\item Sectors were chosen as the first available SPOC product without
checking whether a transit occurs in them, and for long-period planets
most sectors contain none. HD 56414\,b measured 1.82\,h at SNR 1.6 against
a published 7.58\,h.

\item \texttt{context\_days} was removed from the injection configuration
while the generator still referenced it. The injection build crashed on
the first sample.

\item Synthetic periods were drawn from 1\,d upward, below the 2.5-day
extraction window, placing a second transit in the array. Roughly 13\% of
training samples let the network read period from transit spacing rather
than duration.

\item The detectability cut used per-cadence depth over noise rather than
integrated signal-to-noise, admitting transits invisible in the array
whose measured duration was set by noise. Those durations became training
labels.

\item All eight evaluation stars were donating their systematics to the
injection noise pool, which contained only eight distinct stars. Training
and evaluation shared noise, making the injection result uninterpretable
as a test of transfer.

\item The Gaussian likelihood had no floor on $\sigma$, so the model drove
it toward zero on training points. Over 300 epochs the training loss fell
to $-1.42$ while validation loss rose to $+6.07$, with the best epoch at
15.

\item Ephemerides were merged on TIC identifier alone, but multi-planet
systems share a TIC. Five targets carried another planet's transit time:
TOI-201\,c, TOI-2295\,b, HD 28109\,d, TOI-904\,c and HD 22946\,d.

\item Sector candidates were ranked by how centrally the transit sat
rather than by cycle count, ignoring that timing uncertainty compounds
with extrapolation. TOI-2134\,c selected sector 80 ($n=8$) over sector 52
($n=0$), giving depth 0.053\% versus 1.008\% and signal-to-noise 2.5
versus 182.

\end{enumerate}

\end{document}